\documentclass[%
 reprint,
 showkeys,
showpacs,
 amsmath,amssymb,
 aps,
nofootinbib
]{revtex4-1}
\usepackage{graphicx}
\usepackage{dcolumn}
\usepackage{bm}
\usepackage{hyperref}
\usepackage{float}
\usepackage{color}
\usepackage{multirow}
\usepackage{amsmath}
\usepackage{array}

\begin{document}
\preprint{APS/123-QED}
\setcounter{topnumber}{8}
\setcounter{bottomnumber}{8}
\setcounter{totalnumber}{8}

\title{Accurate and Fast reconstruction of Porous Media from Extremely Limited Information Using Conditional Generative Adversarial Network}

\author{Junxi Feng}
 \email{fengjx2011@gmail.com}
\author{Xiaohai He}%
 \email{hxh@scu.edu.cn(Corresponding author)}
\author{Qizhi Teng}%
 \email{qzteng@scu.edu.cn}
\author{Chao Ren}%
\email{chaoren@scu.edu.cn}
\author{Honggang Chen}%
\email{honggang_chen@yeah.net}
\author{Yang Li}%
\email{mongli1989@163.com}
\affiliation{%
 College of Electronics and Information Engineering, Sichuan University, Chengdu 610065, China\\
}%

\date{\today}

\begin{abstract}
Porous media are ubiquitous in both nature and engineering applications, thus their modelling and understanding is of vital importance. In contrast to direct acquisition of three-dimensional (3D) images of such medium, obtaining its sub-region (s) like two-dimensional (2D) images or several small areas could be much feasible. Therefore, reconstructing whole images from the limited information is a primary technique in such cases. Specially, in practice the given data cannot generally be determined by users and may be incomplete or partially informed, thus making existing reconstruction methods inaccurate or even ineffective. To overcome this shortcoming, in this study we proposed a deep learning-based framework for reconstructing full image from its much smaller sub-area(s). Particularly, conditional generative adversarial network (CGAN) is utilized to learn the mapping between input (partial image) and output (full image). To preserve the reconstruction accuracy, two simple but effective objective functions are proposed and then coupled with the other two functions to jointly constrain the training procedure. Due to the inherent essence of this ill-posed problem, a Gaussian noise is introduced for producing reconstruction diversity, thus allowing for providing multiple candidate outputs. Extensively tested on a variety of porous materials and demonstrated by both visual inspection and quantitative comparison, the method is shown to be accurate, stable yet fast ($\sim0.08s$ for a $128 \times 128$  image reconstruction). We highlight that the proposed approach can be readily extended, such as incorporating any user-define conditional data and an arbitrary number of object functions into reconstruction, and being coupled with other reconstruction methods.

\end{abstract}

\maketitle


\section{\label{sec:introduction}Introduction}

Porous media, such as sandstone, soil, alloy, and composite abound in nature and synthetic situations and have play a critical role in a variety of engineering applications. Hence, their understanding and modelling is of significant importance \cite{torquato2013random,sahimi2011flow}.Despite the advance of three-dimensional (3D) imaging techniques like computed tomography (CT) \cite{li2018direct,wang2018three,bostanabad2018computational} and scanning electron microscope (SEM) \cite{tahmasebi2015three}, however, in many cases, only limited data is available for analysis. They could be 3D incomplete data, several two-dimensional (2D) slices and even a few statistics \cite{yeong1998reconstructing,yeong1998reconstructinga}. Therefore, reconstructing 3D full images from these limited data has been a major technique in such situations.

During the past decades, various reconstruction methods have been developed \cite{yeong1998reconstructing,yeong1998reconstructinga,rozman2001efficient,pant2014stochastic,jiao2008modeling,jiao2009superior,chen2015dynamic,
gerke2015improving,gerke2014improving,karsanina2018hierarchical,tang2009pixel,chen2014stable,gao2016pattern,feng2018reconstruction,ju2017multi,
ju20143d,ju20183,okabe2005pore,gao2015reconstruction,ding2018improved,mariethoz2010direct,tahmasebi2012reconstruction,tahmasebi2013cross,
tahmasebi2016enhancing,tahmasebi2016enhancinga,tahmasebi2017hypps,bostanabad2016characterization,bostanabad2016stochastic,feng2018accelerating,mosser2017reconstruction,mosser2018conditioning,mosser2018stochastic,
laloy2017inversion,laloy2018training-image,wang2018porous,li2018markov,li2019super}, and popular algorithms include optimization-based method \cite{yeong1998reconstructing,yeong1998reconstructinga,rozman2001efficient,pant2014stochastic,jiao2008modeling,jiao2009superior,chen2015dynamic,
gerke2015improving,gerke2014improving,karsanina2018hierarchical,tang2009pixel,chen2014stable,gao2016pattern,feng2018reconstruction,ju2017multi,
ju20143d,ju20183}, multi-point statistics (MPS) \cite{okabe2005pore,gao2015reconstruction,ding2018improved}, direct sampling (DS) \cite{mariethoz2010direct}, CCSIM \cite{tahmasebi2012reconstruction,tahmasebi2013cross,tahmasebi2016enhancing,tahmasebi2016enhancinga,tahmasebi2017hypps}, machine learning and deep learning based method \cite{bostanabad2016characterization,bostanabad2016stochastic,feng2018accelerating,mosser2017reconstruction,mosser2018conditioning,mosser2018stochastic,
laloy2017inversion,laloy2018training-image,wang2018porous}, and superdimension method recently proposed \cite{li2018markov,li2019super}. It is well known that generally the prerequisite of this reconstruction methodology is that the 2D training image (TI) needs to meet the requirements of stationarity and ergodicity, in other words, 2D image is able to statistically represent the main characteristic of the entire 3D structure. Despite considerable research on TI selection \cite{mirowski2009stationarity,boisvert2007multiple,gao2017evaluating}, however, in practice the 2D images or data are generally not determined by users, and they may be incomplete or partially applicable \cite{shen2015missing,mariethoz2010reconstruction,sokat2018incomplete}, and thus cannot be directly used as a representative of 3D images. For instance, loss of data or information in 2D images is a universal issue in Earth Sciences and is the primary cause of (hydro) geological uncertainty \cite{mariethoz2010reconstruction}. Clearly, directly reconstructing 3D image from its imperfect 2D images may be infeasible. Instead, to first recover the 2D full image using the limited information and then apply it to 3D reconstruction could be an alternative solution.

Even though 2D image reconstruction can serve as the preparation of the subsequent 3D reconstruction, in fact, reconstruction of 2D image and the analysis based on which can be a distinctive topic. It is noteworthy that practically the 2D images obtained by optical microscope or SEM still play a key role in numerous researches. Notice that in some circumstances acquisition of 2D images could also be tough and highly expensive \cite{ abdollahifard2016improving,ju20183,semnani2017quantifying}. For example, study of nano-scale pore in tight porous materials such as shale generally requires a huge number of high-resolution 2D images for comprehensive analysis, because usually a single imaging field of view (FOV) of such sample is too small to represent the entire material \cite{semnani2017quantifying}. Hence, it is of critical importance and great interest to employ the acquired data (e.g., several small FOVs) for quick reconstruction and accurate investigation, which may save both time and imaging cost.

\begin{figure}[htb]
\includegraphics[keepaspectratio=True, width=230pt]{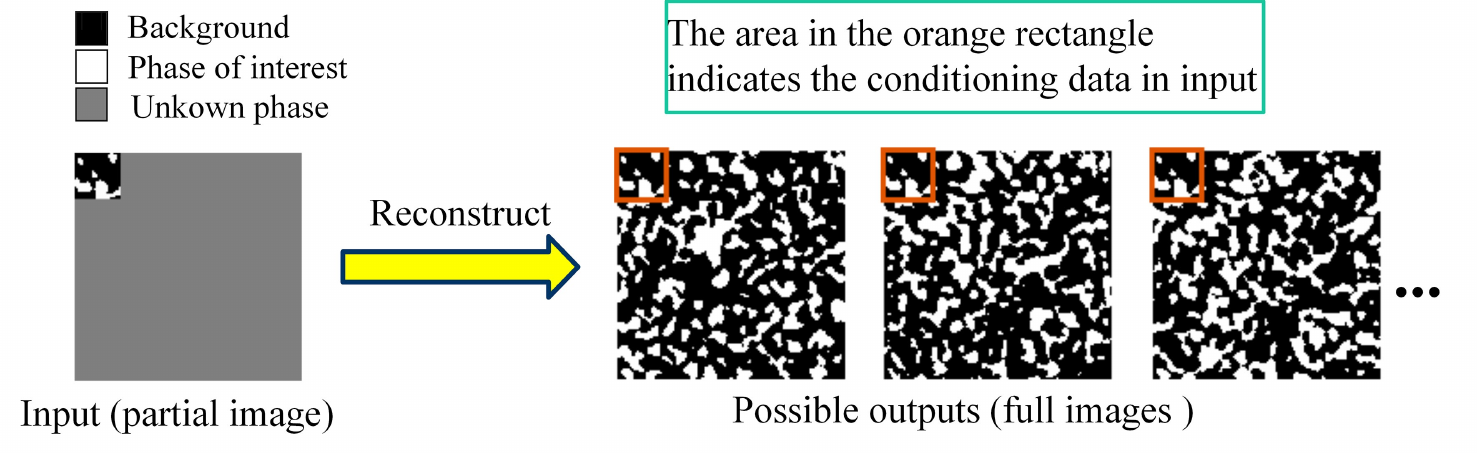}
\caption{Schematic of reconstruction from extremely limited information.}
\label{fig:Fig1_image_transformation}
\end{figure}

Considering the above two aspects, effectively utilizing the given (obtained) data and correspondingly developing accurate 2D/3D reconstruction methods still remain an outstanding problem. Figure \ref{fig:Fig1_image_transformation} shows a schematic of 2D reconstruction of such process. In addition, another related issue is the diversity of reconstruction results. Since the inherent essence of this inverse problem generally enables that more than one solutions are acceptable, the reconstruction algorithm is expected to be not only accurate and fast, but also able to stably provide comparable candidate solutions, thus allowing for user selection.

At present, potential methods for this reconstruction conundrum could be the variants of DS \cite{mariethoz2010direct} and CCSIM \cite{tahmasebi2012reconstruction,tahmasebi2013cross,tahmasebi2016enhancing,tahmasebi2016enhancinga,tahmasebi2017hypps}, and we notice that, the performance of these MPS-like methods heavily rely on the proportion of informed data, i.e., the more data there is, the better performance they may achieve. In the case of extremely limited information here, inaccurate reconstruction and evident artifacts (unnatural structure) may easily arise (see results in Sec.\ref{sec:results_discussion}).

Recently, the advent of machine learning/deep learning techniques has brought new inspirations and insights in a variety of domains \cite{bostanabad2016characterization,bostanabad2016stochastic,lecun2015deep,esteva2017dermatologist,chen2017single,chen2018cisrdcnn,ren2019enhanced,goodfellow2014generative,mirza2014conditional,
karimpouli2019image,karimpouli2019segmentation,tahmasebi2017data,chen2018classification,cang2017microstructure,chan2018exemplar,chan2018parametric,
feng2018accelerating,mosser2017reconstruction,mosser2018conditioning,mosser2018stochastic,laloy2017inversion,laloy2018training-image,wang2018porous,zhu2018multimodal}. Notably, there have been an increasing number of such techniques utilized in the reconstruction and analysis of materials \cite{karimpouli2019image,karimpouli2019segmentation,tahmasebi2017data,chen2018classification,cang2017microstructure,chan2018exemplar,chan2018parametric,
feng2018accelerating,mosser2017reconstruction,mosser2018conditioning,mosser2018stochastic,laloy2017inversion,laloy2018training-image,wang2018porous}. Recent advances include decision tree method \cite{bostanabad2016characterization,bostanabad2016stochastic}, 2D generative adversarial network (GAN) and 2D conditional generative adversarial network(CGAN) \cite{chan2018exemplar,chan2018parametric,feng2018accelerating}, 3D GAN and 3D CGAN \cite{mosser2017reconstruction,mosser2018conditioning,mosser2018stochastic,laloy2017inversion,laloy2018training-image}, and CNN-based method \cite{wang2018porous}. The prevalence of deep learning-based method is due to that by training a neural network using numerous samples (pairs of input and output), it can find a general mapping from input to output. Once neural network is trained, its inference (reconstruction) can be very quick.

In this paper, we aim to propose a deep learning-based framework for reconstructing porous media from extremely limited information. Specially, CGAN is employed to learn a mapping between input (partial image) and output (full image). To preserve the reconstruction accuracy, two simple but effective objective functions (aka, loss functions) are proposed and then coupled with another two loss functions to jointly constrain the training procedure. Additionally, considering the intrinsic nature of this inverse problem that generally more than one reconstructions are reasonable, a Gaussian noise is introduced to produce reconstruction diversity, thus allowing for providing multiple candidate outputs. According to the extensive tests on a variety of porous media and the both visual and quantitative comparisons, our method is demonstrated to be accurate yet stable. Moreover, when given an input, the proposed method can render instant reconstruction on CPU ($\sim0.08s$ for a  $128\times128$ image), which achieves $\sim20$ speedup factor compared with conventional method DS. We remark that apart from the ability to solve the problem of incomplete data, our approach can be readily extended, such as incorporating an arbitrary number of object functions of any types into reconstruction, the ability to incorporate any user-define conditional data, and being coupled with other reconstruction methods.

The rest of this paper is organized as follows: Section \ref{sec:cgan_reconstruction} details the reconstruction framework of porous media, including fundamental of CGAN, introduction of noise, and the loss functions. Section \ref{sec:reconstruction_assessment} describes the assessment methods for reconstruction. Results and comparisons are demonstrated in Sec. \ref{sec:results_discussion}. In Section \ref{sec:conclusion}, we make concluding remarks.

\begin{figure*}[!htb]
\includegraphics[keepaspectratio=True, width=430pt]{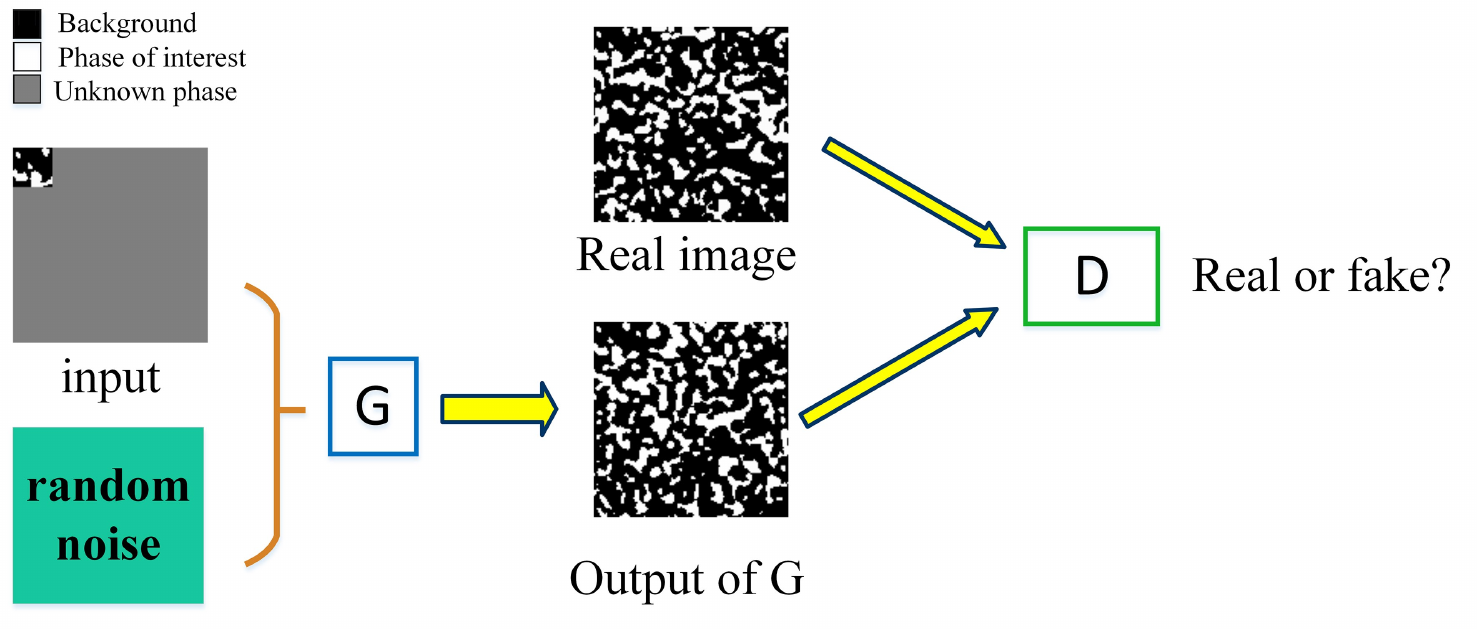}
\caption{\label{fig:Fig2_schematic_of_cgan}Schematic of CGAN. $G$ tries to generate realistic data from given inputs to fool $D$, while $D$ attempts to identify the fakes by $G$ and real data.}
\end{figure*}

\section{\label{sec:cgan_reconstruction}Reconstruction of porous media using conditional generative adversarial network}

In this section, we will present the details on the reconstruction of porous media using conditional generative adversarial network (CGAN), including primary principle of CGAN, design of network architectures, and loss functions.

\subsection{\label{sec:cgan_principle}Principle of CGAN}
Owing to CGAN \cite{mirza2014conditional} is the conditional version of GAN \cite{goodfellow2014generative}, we would like to first introduce GAN. In particular, GAN consists of two ¡®adversarial¡¯ sub-networks, one is generator $G$ and the other is discriminator $D$. Especially, they battle in a two-player min-max game in which the generator $G$ tries to generate realistic data from given input to fool discriminator $D$, whereas $D$ attempts to identify the fakes by $G$ and real data (target). The goal of $G$ is to learn a mapping $G(\mathbf{z})$ that maps prior noise ${p_z}\left(\mathbf{z} \right)$ to the real data $\mathbf{y}$, while the output of discriminator $D(\mathbf{y})$ is to give the probability that $\mathbf{y}$ is from training data rather than the fake data generated by $G$. Both $G$ and $D$ are trained alternately to optimize the following objective function:

\begin{equation}
\begin{aligned}
\mathop {\min }\limits_G \mathop {\max }\limits_D {\rm{ }}{L_{GAN}}\left( {G,D} \right) = {E_{\mathbf{y} \sim {p_{data}}\left( \mathbf{y} \right)}}\left[ {\log D\left( \mathbf{y} \right)} \right]\\ + {E_{\mathbf{z} \sim {p_\mathbf{z}}\left( \mathbf{z} \right)}}\left[ {\log \left( {1 - D\left( {G\left( \mathbf{z} \right)} \right)} \right)} \right]
\label{equ:Eq1_gan_loss}
\end{aligned}
\end{equation}

where $G$ tries to minimize this expression and in the same time $D$ tries to maximize it, thus indicating the ¡®adversarial¡¯ conception in GAN.
Notably, at the very beginning, the abilities of both $G$ and $D$ are quite weak; with iterations, they gradually evolve to be powerful and finally reach Nash equilibrium \cite{goodfellow2014generative}, in which $G$ is able to produce realistic data which cannot be recognized by $D$. Once trained, $D$ is discarded and only $G$ is used to transform input to its expected output. In general, the use of initial GAN is to generate realistic data using noise distribution such as Gaussian or rand noise from the given samples, which could be of special use in the situations that data is lacking.

As an improved version of GAN, the CGAN allows for incorporating conditional data as an external input $\mathbf{x}$, rather than only a noise distribution $\mathbf{z}$, as demonstrated in Fig.\ref{fig:Fig2_schematic_of_cgan}.

Therefore, the modified objective function is given as:
\begin{small}
\begin{equation}
\begin{aligned}
\mathop {\min }\limits_G \mathop {\max }\limits_D {L_{CGAN}}\left( {D,G} \right) = {E_{\mathbf{x} \sim {p_{data}}\left( \mathbf{x} \right),\mathbf{y} \sim {p_{data}}\left( \mathbf{y} \right)}}\left[ {\log D\left( {\mathbf{x},\mathbf{y}} \right)} \right]\\ + {E_{\mathbf{x} \sim {p_{data}}\left( \mathbf{x} \right),\mathbf{z} \sim {p_\mathbf{z}}\left( \mathbf{z} \right)}}\left[ {\log \left( {1 - D\left( {\mathbf{x},G\left( {\mathbf{x},\mathbf{z}} \right)} \right)} \right)} \right]
\label{equ:Eq2_cgan_loss}
\end{aligned}
\end{equation}
\end{small}

\begin{figure*}[!htb]
\includegraphics[keepaspectratio=True, width=\textwidth]{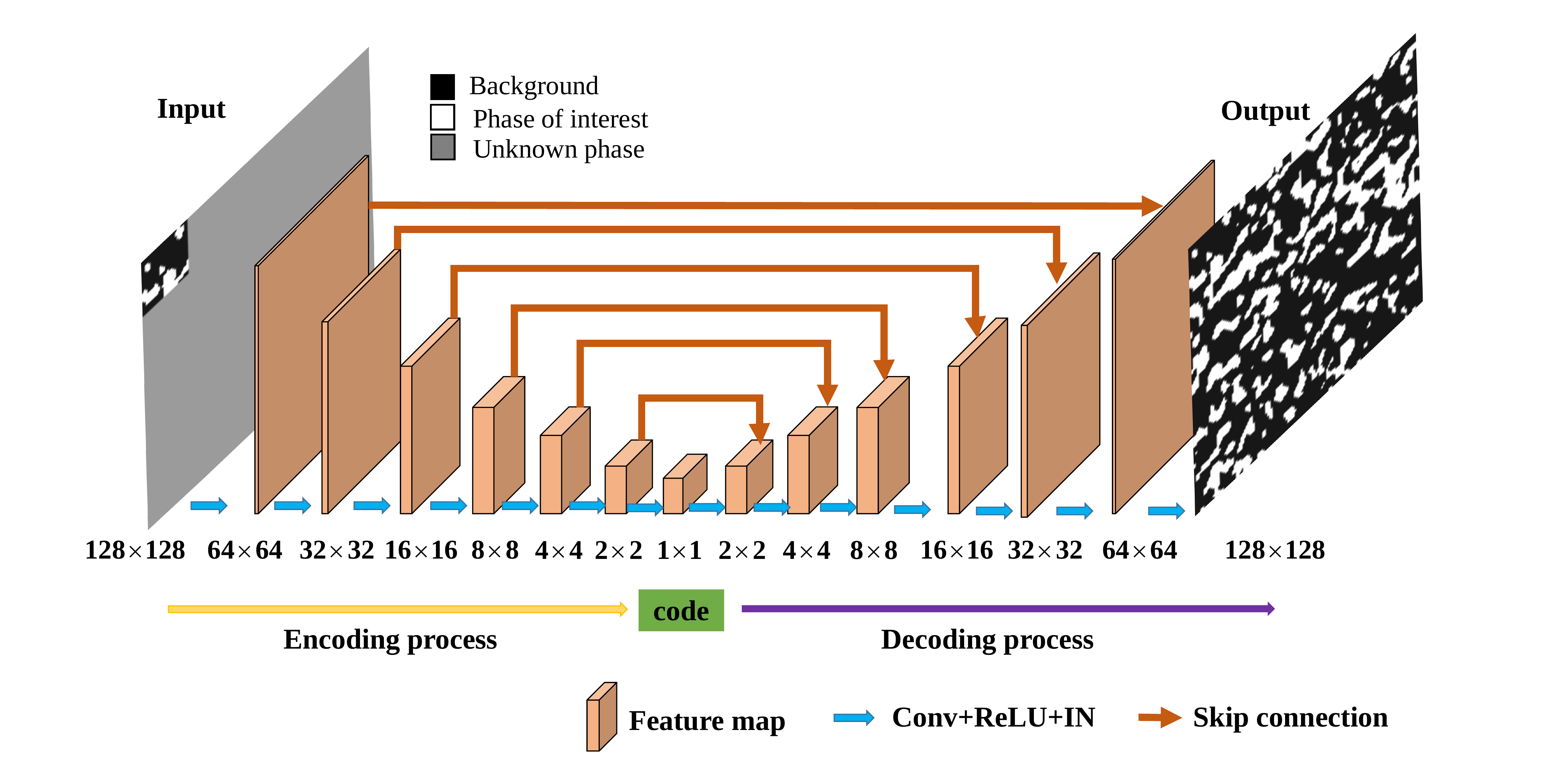}
\caption{Schematic of architecture of network G.}
\label{fig:Fig3_architecture_of_g}
\end{figure*}

Actually, in terms of different tasks, the noise in CGAN could be added or not. Especially, the noise could even be dropped in the case of one-to-one image translation problems \cite{isola2017image}. However, in this paper, the diversity of reconstruction is mainly due to the introduction of noise. Hence, the noise is reserved and fused together with $G$ (see more details presented in subsection \ref{sec:noise_injection}).

\subsection{\label{sec:network_architectures}Network architectures}

In the following, we will elaborate the architectures of generator G and discriminator D, as well as the injection of noise into G.

\subsubsection{\label{sec:G_architecture}Architecture of generator}

Specially, in general image-to-image tasks, U-Net network architecture is frequently employed, for its merit that relatively fewer parameters and multi-scale characteristic for feature extraction. Owing to the nature of reconstruction of porous media is also an image-to-image task, we mainly follow the design of BicycleGAN \cite{zhu2018multimodal}, as it is a representative U-Net based framework. Figure \ref{fig:Fig3_architecture_of_g} depicts the main architecture of G. Specifically, this network performs two main steps: encoding process by encoder and decoding process by decoder. These two procedures could be viewed as the nonlinear down-sampling and up-sampling of input image. In this work we focus on $128 \times 128$ images, so starting from the size of $128 \times 128$, the input image is convoluted and down-sampled gradually to $1 \times 1$ code by the sampling factor 2, and then deconvoluted and up-sampled back to $128 \times 128$ for output by the same factor 2. Each convolutional layer or deconvolutional is followed by nonlinear activation layer (ReLU or Leaky ReLU) and instance normalization (IN) layer. In general, the encoding and decoding processes of U-Net are symmetric, which means that the shape of feature maps in the symmetric position are the same. Thus, a skip connection is usually added to assist decoding process by introducing the feature map information of encoding process.

\subsubsection{\label{sec:D_architecture}Architecture of discriminator}

\begin{figure*}[htb]
\includegraphics[keepaspectratio=True, width=430pt]{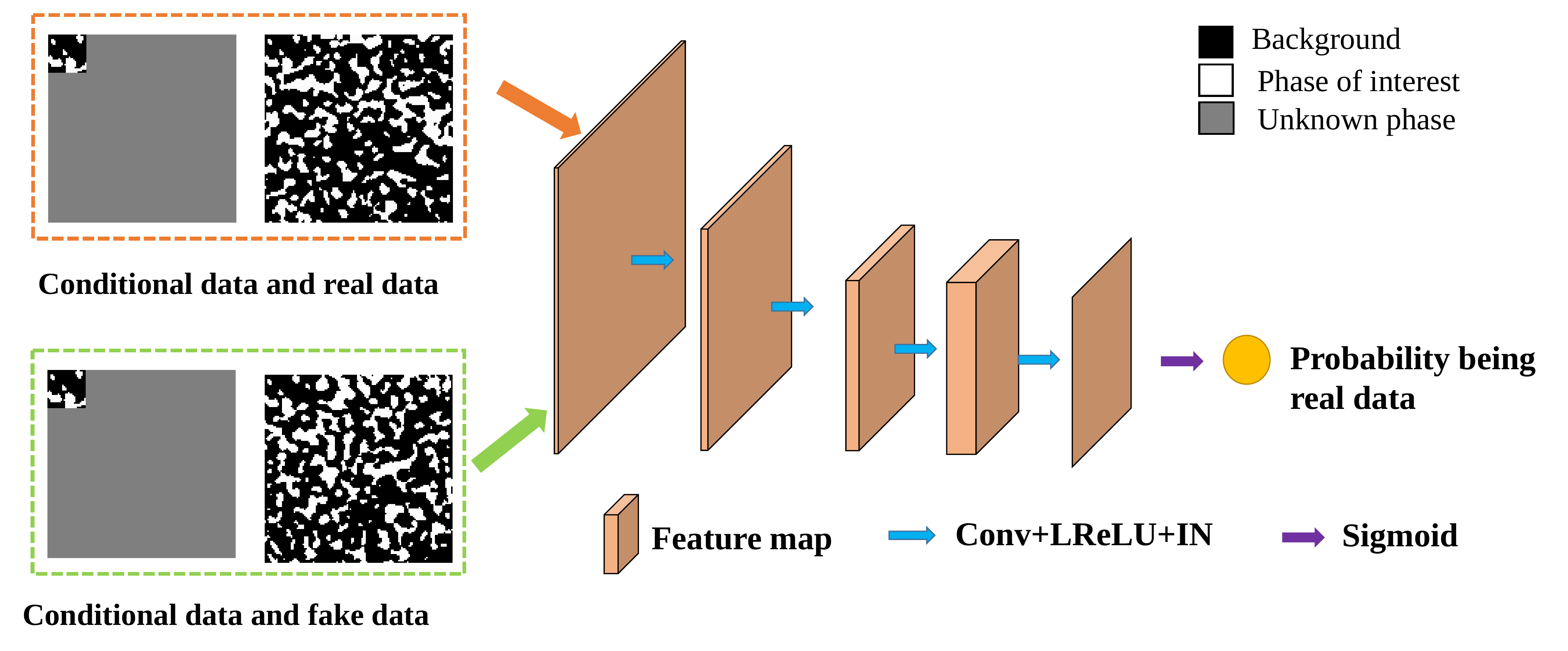}
\caption{Schematic of architecture of network D.}
\label{fig:Fig4_architecture_of_d}
\end{figure*}

The design of discriminator $D$ is relatively simpler than generator G, as shown in Fig. \ref{fig:Fig4_architecture_of_d}. It is composed of five convolutional layers, each of them is followed by Leaky ReLU (LReLU) layer and instance normalization (IN) layer. As aforementioned, $D$ is trained to distinguish the real data $\mathbf{y}$ and fake data $G(\mathbf{z})$. Hence, the input of $D$ is the pair of conditional data $\mathbf{x}$ and real data $\mathbf{y}$, or that of conditional data $\mathbf{x}$ and fake data $G(\mathbf{z})$ during training process. By using the sigmoid function and an additional average function, the output of $D$ is transformed to a probability between $0$ to $1$, which indicates how real the data is. The D is trained to recognize real data $\mathbf{y}$ as probability $1$ and fake data $G(\mathbf{z})$ as probability $0$. When reaching the final balance with $G$, both real and fake data are identified as probability $0.5$.

\subsubsection{\label{sec:noise_injection}Noise injection}

\begin{figure*}[htb]
\includegraphics[keepaspectratio=True, width=430pt]{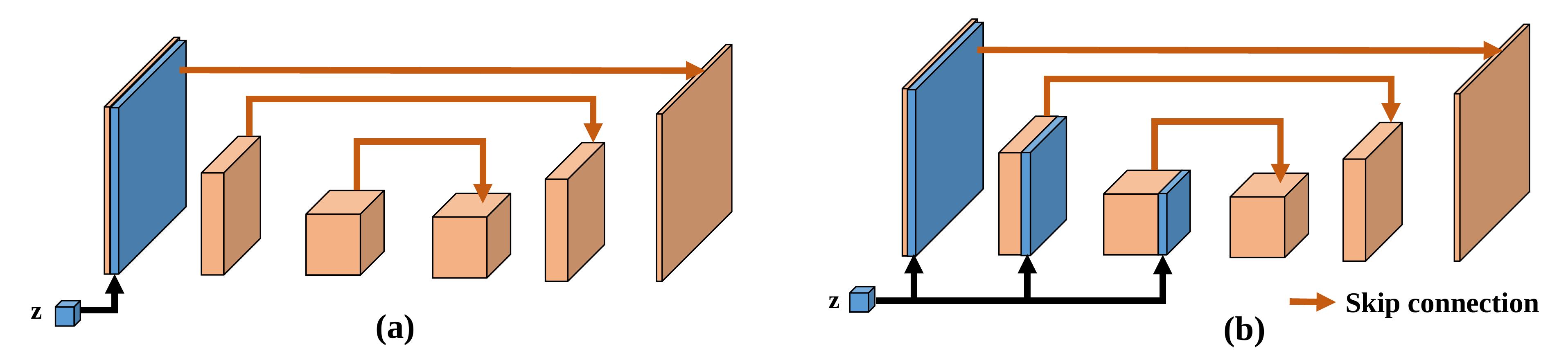}
\caption{\label{fig:Fig5_noise_injection} Alternatives for noise injection. Noise $\mathbf{z}$ is injected by spatial replication and concatenation into the generator. (a) Adding noise in the first layer and (b) in each layer of the encoder.}
\end{figure*}

It is worth noting that the goal of injection of random noise into generator $G$ is to introduce its output diversity, for providing multiple choices for users due to the inherent essence of this ill-posed problem. In general, there are two alternatives of noise injection: i) adding noise in the first layer (Fig. \ref{fig:Fig5_noise_injection} a) and ii) in each layer of the encoder (Fig. \ref{fig:Fig5_noise_injection} b). As demonstrated in the literature \cite{zhu2018multimodal}, the noise injected all layers in the encoder leads to a slightly better performance. Hence, in the proposed method we utilize the design in Fig. \ref{fig:Fig5_noise_injection} b.

In our work, we use a Gaussian noise with a shape $n_z \times 1 \times 1$, which is first spatially replicated to same height and width of the current layer of encoder, and then is concatenated with this layer along channel dimension. For instance, assume the shape of current layer is $C \times H \times W$, the Gaussian noise $n_z \times 1 \times 1$ is spatially replicated to $n_z \times H \times W$ along H and W dimension. After concatenation along channel dimension, the shape of resulting concatenated layer will be $(C+n_z) \times H \times W$.

\subsection{\label{sec:loss}Loss function}

Similar to other optimization-based approaches like simulated annealing \cite{yeong1998reconstructing,yeong1998reconstructinga,rozman2001efficient,pant2014stochastic,jiao2008modeling,jiao2009superior,chen2015dynamic,
gerke2015improving,gerke2014improving,karsanina2018hierarchical,tang2009pixel,chen2014stable,gao2016pattern,feng2018reconstruction,ju2017multi,
ju20143d,ju20183}, the goal of deep learning-based method is to minimize the loss function, which is used to evaluate the discrepancy between network prediction and target. On the basis of Eq. \ref{equ:Eq2_cgan_loss}, the loss of discriminator D can be equivalently concluded to minimize the equation as follows:

\begin{equation}
{L_D}=  - \left\{ \begin{array}{l}
{E_{\mathbf{x} \sim {p_{data}}\left( \mathbf{x} \right),\mathbf{y} \sim {p_{data}}\left( \mathbf{y} \right)}}\left[ {\log D\left( {\mathbf{x},\mathbf{y}} \right)} \right] + \\
{E_{\mathbf{x} \sim {p_{data}}\left( \mathbf{x} \right),\mathbf{z} \sim {p_\mathbf{z}}\left( \mathbf{z} \right)}}\left[ {\log \left( {1 - D\left( {\mathbf{x},G\left( {\mathbf{x},\mathbf{z}} \right)} \right)} \right)} \right]
\end{array} \right\}
\label{equ:Eq3_cgan_loss_d}
\end{equation}

In what follows, we would like to detail the loss function of generator G, since it significantly determines the quality of generated results. Specially, the total loss function of $G$ is composed of four individual loss functions, namely, CGAN loss $L_G$, L1 loss $L_{L1}$, and two proposed loss functions for this reconstruction task, i.e., pattern loss $L_{pattern}$ and porosity loss $L_{posority}$.

CGAN loss $L_G$ comes from the fundamental of CGAN framework, representing how close output of $G$ is to the target when discriminated by $D$ \cite{goodfellow2014generative}. In terms of Equation Eq. \ref{equ:Eq2_cgan_loss}, the loss $L_G$ is defined as:

\begin{equation}
{L_G} = {E_{\mathbf{x} \sim {p_{data}}\left( \mathbf{x} \right),\mathbf{z} \sim {p_z}\left( \mathbf{z} \right)}}\left[ {\log \left( {1 - D\left( {\mathbf{x},G\left( {\mathbf{x},\mathbf{z}} \right)} \right)} \right)} \right]
\label{equ:Eq4_cgan_loss_g}
\end{equation}

L1 loss $L_{L1}$, defined as a sum of pixel-wise absolute value difference between $G's$ output $G(\mathbf{x},\mathbf{z})$ and input conditional data $\mathbf{x}$, is given by:

\begin{equation}
{L_{{L_1}}} = {\left\| {\mathbf{x} - G\left( {\mathbf{x},\mathbf{z}} \right)} \right\|_1},
\label{equ:Eq5_cgan_L1_loss}
\end{equation}
\\
where $\mathbf{z}$ is Gaussian noise. This loss function is utilized to ensure that the output $G(\mathbf{x},\mathbf{z})$ keeps the same conditioning data contained in $\mathbf{x}$, including both value and position.

The third one is pattern loss $L_{pattern}$, proposed to quantify the mean squared error (MSE) of pattern distributions in the respective images of $G's$ output $G(\mathbf{x},\mathbf{z})$ and target $\mathbf{y}$. This loss function is leveraged to describe the local texture difference in two images, and the definition is:

\begin{equation}
{L_{pattern}} = \left\| {{y_{pattern}} - {{\left( {\mathbf{G}\left( {\mathbf{x},\mathbf{z}} \right)} \right)}_{pattern}}} \right\|_2^2
\label{equ:Eq6_cgan_pattern_loss}
\end{equation}

\begin{figure*}[htb]
\includegraphics[keepaspectratio=True, width=430pt]{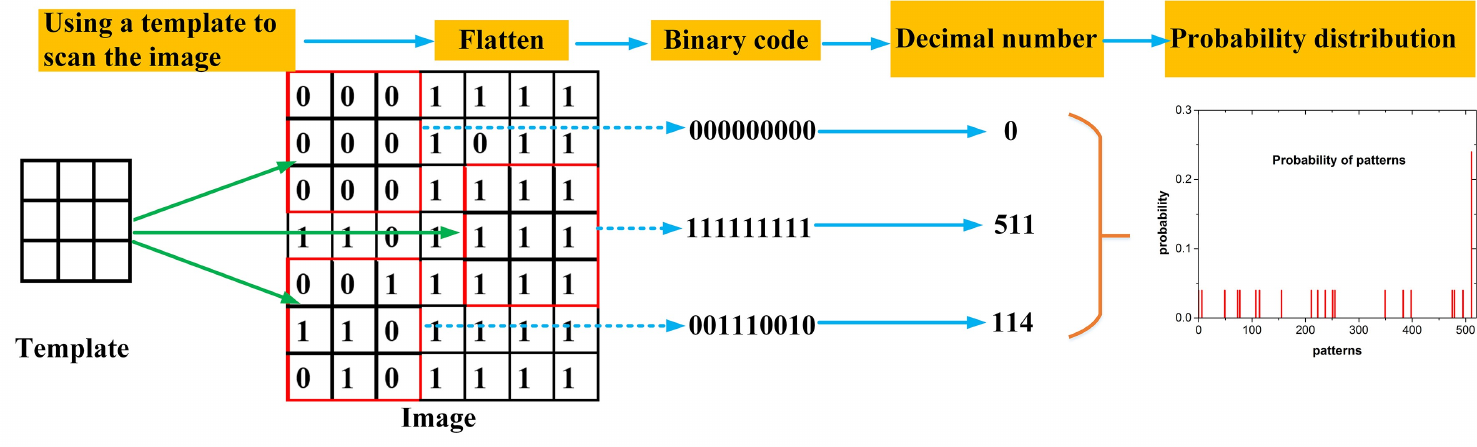}
\caption{\label{fig:Fig6_Pattern_loss}Schematic of obtainment of probability distribution of patterns.}
\end{figure*}

Particularly, the pattern in an image is defined as the multi-point configuration captured by a template. The calculation of pattern loss is straightforward: i) by using a $N \times N$ template, scan the image and collect all the patterns; ii) flatten each pattern, and obtain its corresponding binary code and then convert it to decimal number; iii) the occurrence of each decimal number is counted and then is normalized (divided by the total number of patterns in the image), and thus the probability distribution of patterns is obtained; iv) calculate the Euclidean distance between the two distributions of target and reconstruction. Figure \ref{fig:Fig6_Pattern_loss} gives the main steps of such process.

The last loss function is porosity loss $L_{posority}$, which is presented to identify the disagreement of porosity of output $G(\mathbf{x},\mathbf{z})$ and target $\mathbf{y}$, and thus maintaining the porosity agreement during reconstruction. It is given as follows:

\begin{equation}
{L_{porosity}} = \left\| {{y_{porosity}} - {{\left( {G\left( {\mathbf{x},\mathbf{z}} \right)} \right)}_{porosity}}} \right\|_2^2
\label{equ:Eq7_cgan_porosity_loss}
\end{equation}

The total loss is a weighted sum of above four loss functions and is defined as:

\begin{equation}
\begin{aligned}
\begin{array}{c}
{L_{total{\rm{ }}loss}} = {L_{CGAN}}\left( {G,D} \right) + {\lambda _{{L_1}}}{L_{{L_1}}}\\
 + {\lambda _{pattern}}{L_{pattern}} + {\lambda _{porosity}}{L_{porosity}},
\end{array}
\label{equ:Eq8_cgan_total_loss}
\end{aligned}
\end{equation}
\\
where the hyper-parameters $\lambda _{{L_1}}$, $\lambda _{pattern}$ and $\lambda _{porosity}$ control the relative importance of each term.

\section{\label{sec:reconstruction_assessment}Assessment methods for reconstruction}

\begin{figure}[!htb]
\includegraphics[keepaspectratio=True, width=210pt]{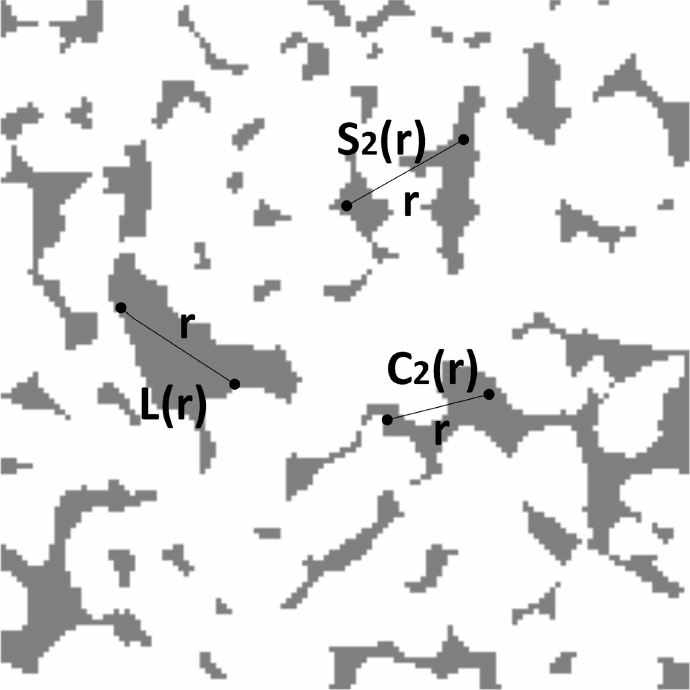}
\caption{\label{fig:Fig7_Three_morphological_functions}Definition of three morphological functions.}
\end{figure}

In this paper, to verify the performance of the proposed method, porosity and three morphological functions, namely, two-point correlation function \cite{torquato2013random}, lineal-path function \cite{lu1992lineal} and two-point cluster function \cite{torquato1988two-point}, are employed. Figure \ref{fig:Fig7_Three_morphological_functions} illustrates the definition of these functions.

Two-point correlation function $S_2(\mathbf{r})$ gives the probability that two points separated by a distance $r=\|\mathbf{r}\|$ both lie in the same phase, indicating spatial correlation of two points. For statistically isotropic and homogeneous media, it only relies on the distance   between the point pair. Therefore, for briefness $r$ is dropped.

Similar to the $S_2$, the two-point cluster function $C_2$  presents the probability that two points both lie in the same cluster. In terms of its definition, it embodies higher-order morphological information than $S_2$.

The other descriptor is lineal-path function $L$, which is the probability that a segment entirely lies in the same phase. It encodes connectedness information of medium along a lineal path.

\section{\label{sec:results_discussion}Results and discussion}

Here, we focus on 2D two-phase structures, while with slight modifications, extensions of multi-phase reconstruction and 3D reconstruction are also straightforward. To ascertain the performance of our method CGAN, we test it on four types of porous media, covering high or low porosity, and isotropy or anisotropy.

\subsection{\label{sec:dataset}Dataset}

\begin{figure}[!htb]
\includegraphics[keepaspectratio=True, width=250pt]{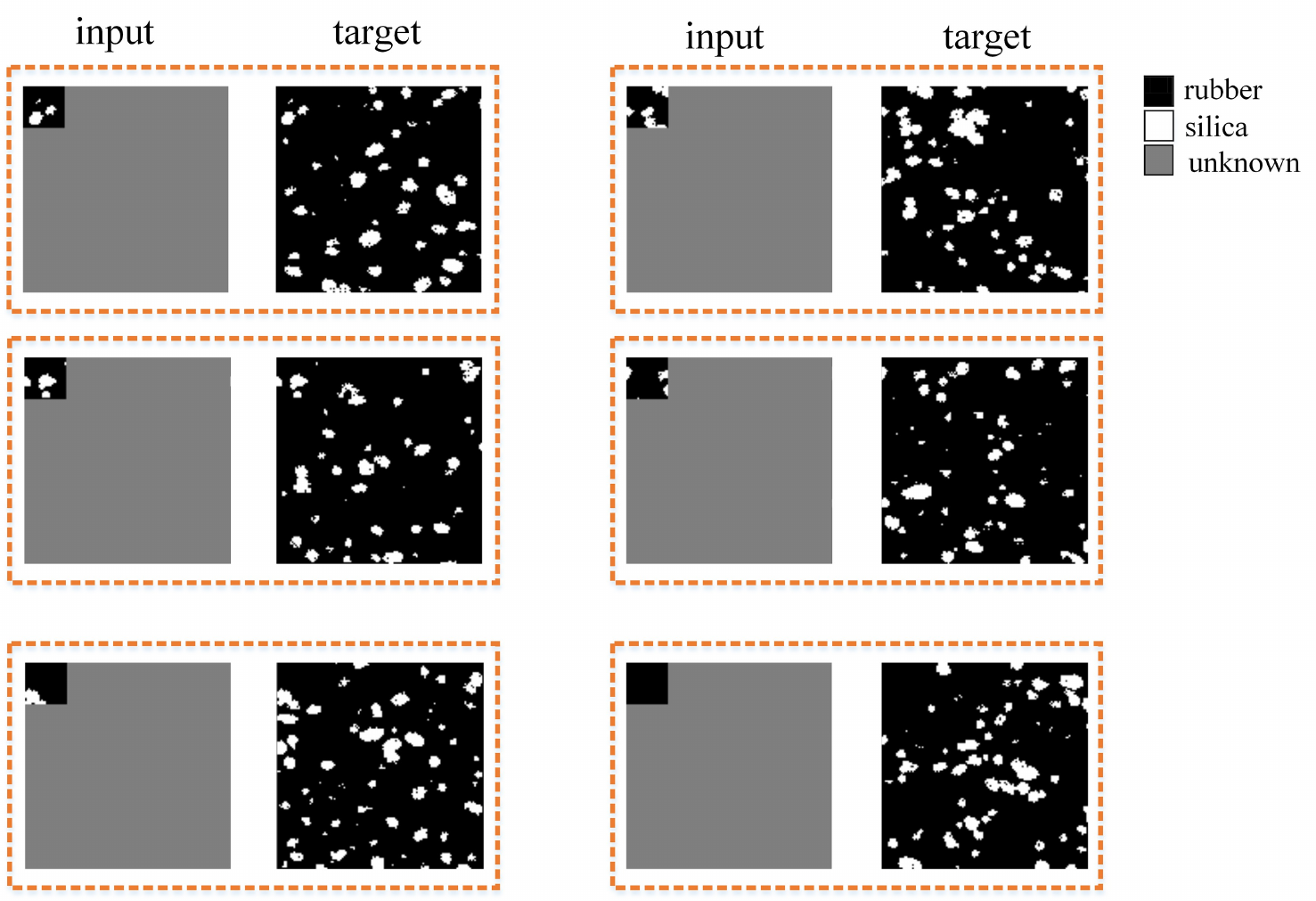}
\caption{\label{fig:Fig8_samples_in_silica_material_dataset}Six samples in silica material dataset.}
\end{figure}

Notice that in this work, in terms of each category of porous medium, an associated dataset is made, which encompasses 600-1000 images. Especially, 70$\%$ samples in each dataset is randomly chosen as training set and the rest 30$\%$ is as testing set. Figure \ref{fig:Fig8_samples_in_silica_material_dataset} presents six examples in the dataset of silica material, each sample is a pair of input (partial image) and target (full image).

In all of our reconstruction tasks here, hyper-parameter $\lambda _{{L_1}}$, $\lambda _{pattern}$ and $\lambda _{porosity}$ in Eq. \ref{equ:Eq8_cgan_total_loss} are empirically set to $10$,  $5.0 \times 10^5$ and $1.0 \times 10^3$, respectively. For the tradeoff between accuracy and efficiency of pattern loss, template size $N$ in calculation of pattern loss is set to $3$. In addition, batch size and channel $n_z$ of noise are respectively set to $2$ and $8$. For both $G$ and $D$, we use Adam optimizer \cite{kingma2014adam} with an initial learning rate $0.0002$, and a linear decay for stable training.

\subsection{\label{sec:results_comparisons}Results and comparisons}

In this subsection, we present reconstruction results and comparisons of four porous media, including isotropic materials such as silica, battery material, sandstone, and an anisotropic medium. In particular, porosity and three morphological functions, i.e., $S_2$, $L$, and $C_2$ are employed to evaluate the reconstruction accuracy. For each medium, $20$ realizations are generated and an additional average of statistics (morphological functions and porosity) over them is also presented.

\subsubsection{\label{sec:silica_reconstruction}Silica reconstruction}

\begin{figure*}[!htb]
\includegraphics[keepaspectratio=True, width=430pt]{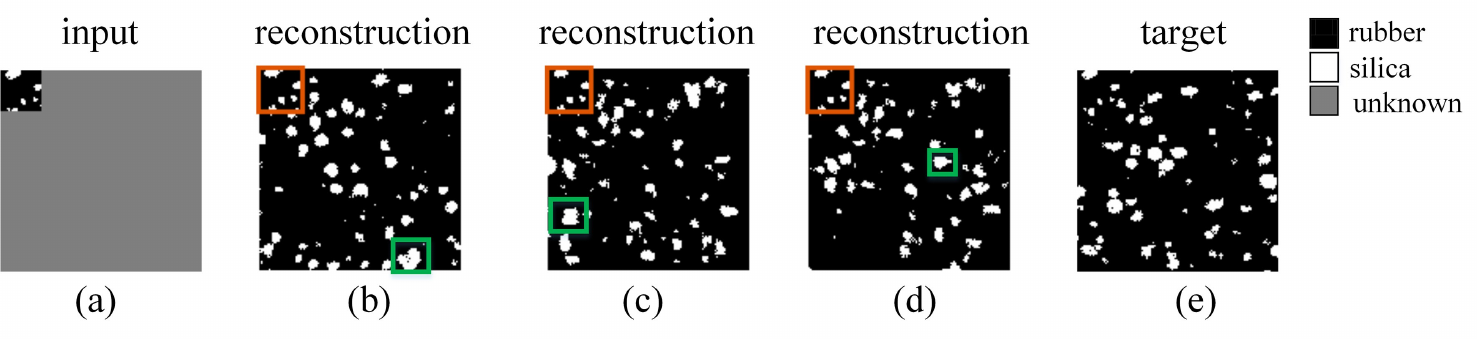}
\caption{\label{fig:Fig9_silica_reconstruction}Input (a), a $26 \times 26 $ square in the top-left region of the target (e), and three realizations of CGAN (b)-(d). Orange rectangles show the reproduction of hard data in the input, while green rectangles indicate the much bigger white cluster in reconstructions than those in the input, which demonstrates the introduction of external data by our method.}
\end{figure*}

First, we reconstructed an isotropic porous material, i.e., silica in a rubber matrix with a size of  $128 \times 128 $ and a low porosity ${\phi _{target}} = 0.0962$. Figure \ref{fig:Fig9_silica_reconstruction} demonstrates the input, three randomly selected reconstructions, as well as the target. Notice, the informed data in the input is a $26 \times 26 $ square in the top-left area of the target, only accounting for $\sim 4\%$ data in this image. However, the reconstructions are still visually indistinguishable from the target. It can be obviously seen that, the hard data in the input is well honored in all of the reconstructions (marked in orange rectangles), while the rest of the image keeps good diversity, which allows for multiple selections for users.

\begin{figure*}[!htb]
\includegraphics[keepaspectratio=True, width=430pt]{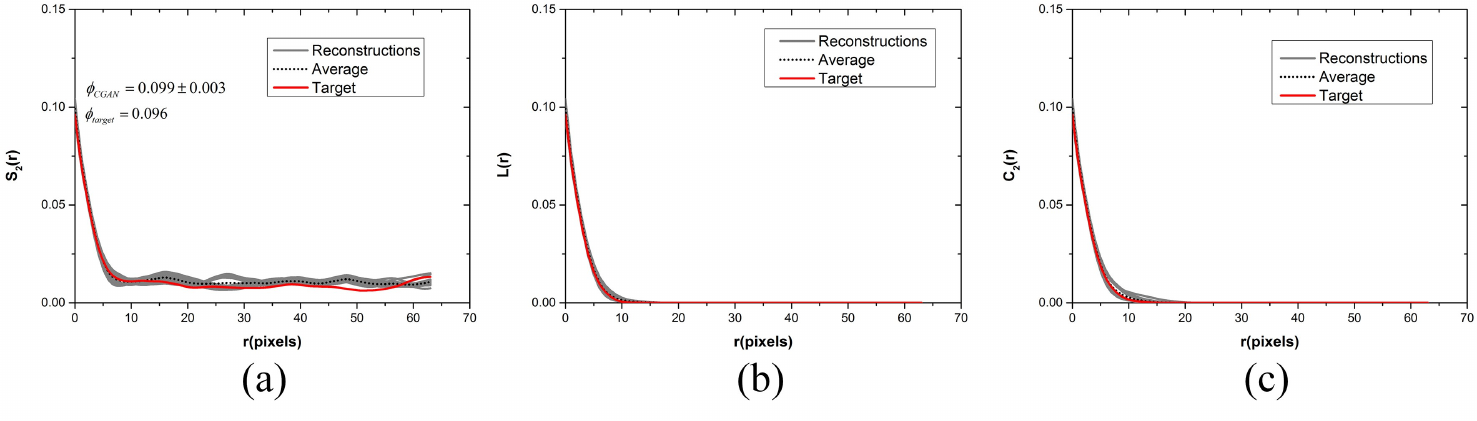}
\caption{\label{fig:Fig10_silica_comparison}Comparison of statistical functions between $20$ reconstructions, their average, and target. The calculation of statistical functions is along X and Y directions and then averaged.}
\end{figure*}

Moreover, in addition to visual inspection, we further quantitatively compare the reconstruction accuracy of the proposed method. Figure \ref{fig:Fig10_silica_comparison} depicts the quantitative comparison of statistical functions and porosities between $20$ reconstructions, their average, and the target. Good agreement can be observed between reconstructions and target, and the average over $20$ realizations excellently matches the target, demonstrating the accuracy of proposed method. Additionally, the small biases of functions between reconstructions and target also indicates the stability of our method. The porosity distribution (Fig. \ref{fig:Fig10_silica_comparison} a) on $20$ reconstructions is ${\phi _{CGAN}} = 0.099 \pm 0.003$ $(Mean \pm Standard \; Deviation)$ while the target is ${\phi _{target}} = 0.0962 $, their accordance also manifests the accuracy and robustness of the proposed method.

Notably, once trained, our method only takes $\sim0.08s$ for reconstruction when running on an Intel i7-4790K ( $4.00$ GHz) CPU. It is also worth noting that this input image (Fig. \ref{fig:Fig9_silica_reconstruction} a) may be regarded as a high-resolution image with small FOV and by using our method, its much larger FOV with high resolution can be accurately recovered. Meanwhile, according to the input and reconstructions (Fig. \ref{fig:Fig9_silica_reconstruction}), some of white clusters (marked in green rectangles) in reconstruction are much bigger than the given ones, therefore indicating the additional information is indeed introduced by our method.

\subsubsection{\label{sec:battery_reconstruction}Battery material reconstruction}

\begin{figure*}[!htb]
\includegraphics[keepaspectratio=True, width=430pt]{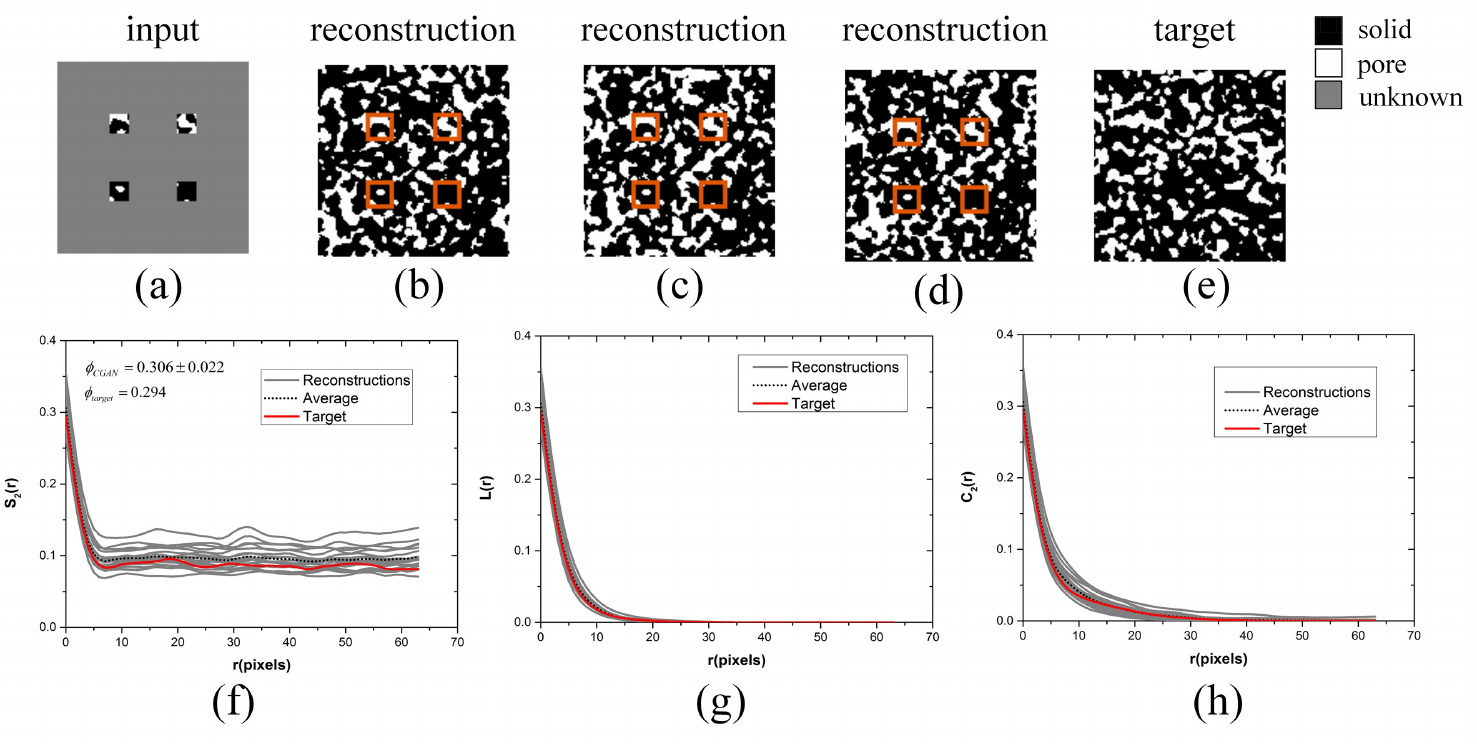}
\caption{\label{fig:Fig11_battery_comparison}Comparison of visual inspection and statistical functions. Orange rectangles show the reproduction of hard data. The calculation of statistical functions is along X and Y directions and then averaged.}
\end{figure*}

In addition to one subarea reconstruction, our method is also verified on an isotropic battery material \cite{ananyev2018degradation} with four tiny $13 \times 13 $ subareas informed and target porosity ${\phi _{target}} = 0.294$, as shown in Fig. \ref{fig:Fig11_battery_comparison} a and Fig.\ref{fig:Fig11_battery_comparison} e. The purpose of this experiment is to mimic the extreme circumstances in aerospace \cite{shen2015missing}, geoscience \cite{mariethoz2010reconstruction,sokat2018incomplete}, etc., that only several incomplete parts of data may be available. Again, according to the comparisons in Fig. \ref{fig:Fig11_battery_comparison}, the reconstructions are visually comparable, and in terms of both porosity (${\phi _{CGAN}} = 0.306 \pm 0.022$ ) and statistical functions, the agreement is excellent.

We would like to emphasize that the amount of informed data in the input image of this material is the same as that of silica, however, here the standard deviation $0.022$ of its reconstruction porosities is much higher, in comparison with that of the silica reconstruction ($0.003$). This phenomenon can be also be evidently observed from the larger vibrations of the three descriptors in Fig. \ref{fig:Fig11_battery_comparison}(f)-(h). Actually, this is expected since this material is more complex in both connection and geometry than those of silica material.

\subsubsection{\label{sec:sandstone_reconstruction}Sandstone reconstruction}

\begin{figure*}[!htb]
\includegraphics[keepaspectratio=True, width=430pt]{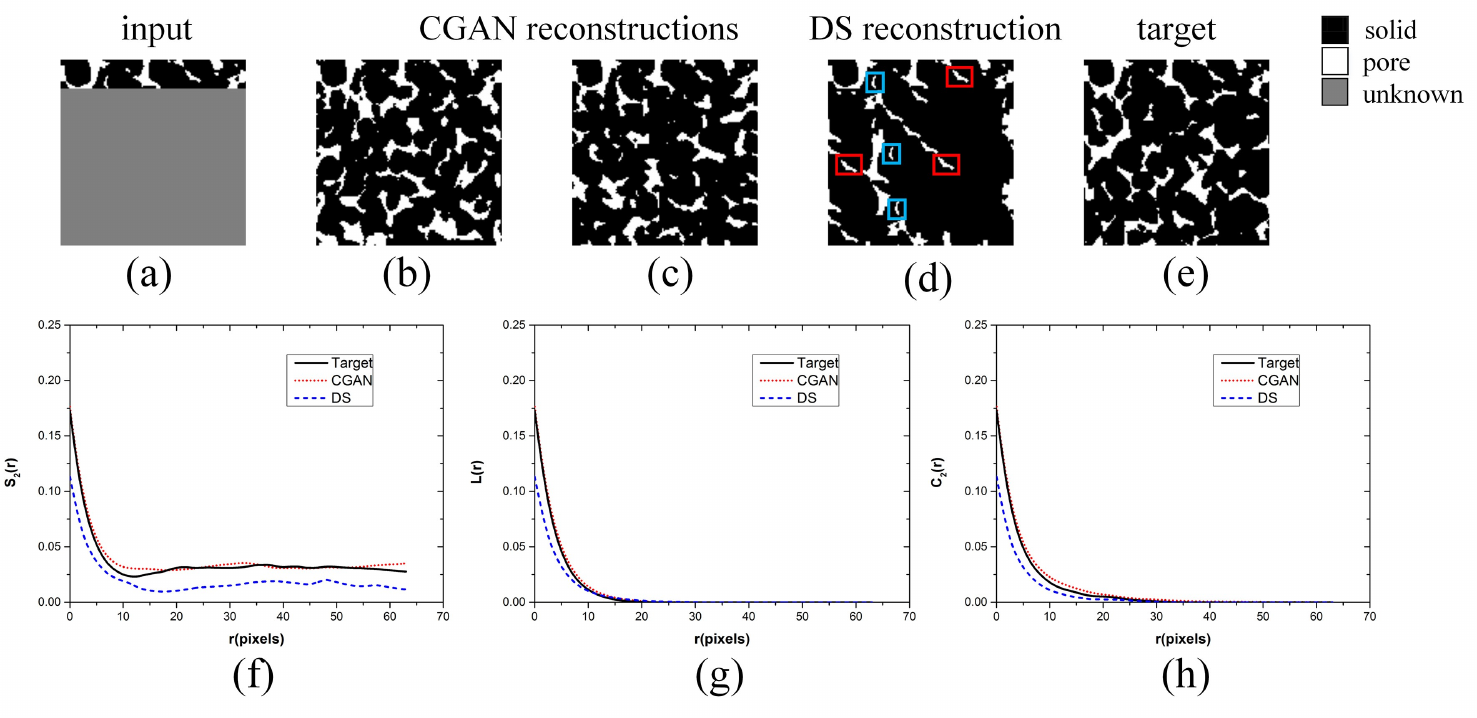}
\caption{\label{fig:Fig12_sandstone_comparison}Comparison of visual inspection and statistical functions. Red and blue rectangles respectively present the repeat of two patterns of hard data in input. The calculation of statistical functions is along X and Y directions and then averaged.}
\end{figure*}

It is useful to compare the performances of our method CGAN and another existing method, i.e., a variation of $DS$ (here also called $DS$ for convenience). As presented in Fig. \ref{fig:Fig12_sandstone_comparison} a, we use a sandstone image with a relatively larger region ($128 \times 20$) of known data for the demonstration purpose. Obviously, In light of both visual and quantitative comparisons in Fig. \ref{fig:Fig12_sandstone_comparison}, the performance of CGAN significantly surpasses that of DS. As can be observed in Fig. \ref{fig:Fig12_sandstone_comparison} d, the reconstruction by DS is unnatural, in which the pore size is severely underestimated and the connectedness is also poor, compared with the target image. Specially, the reasons for this are twofold: i) the reconstruction mechanism of sequential simulation of DS can readily cause error accumulation and thus make reconstruction inaccurate; ii) more importantly, the nature of this method is generally directly repeating the given data (usually in the manner of patterns) into the rest of unknown part, rather than generating or introducing additional realistic information. As can be seen Fig. (\ref{fig:Fig12_sandstone_comparison} d), rectangles marked in red and blue demonstrate the repeat of two patterns of hard data in input.

\begin{figure*}[!htb]
\includegraphics[keepaspectratio=True, width=430pt]{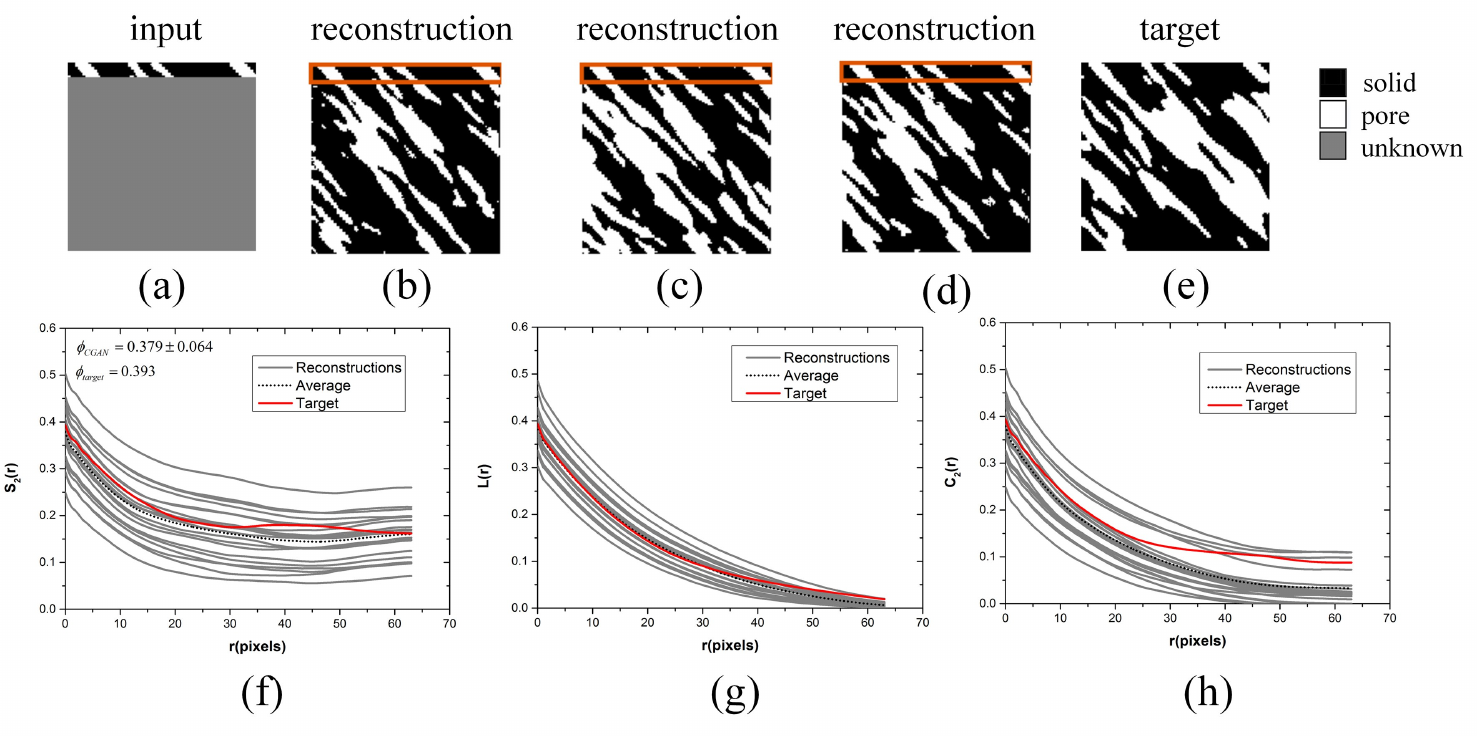}
\caption{\label{fig:Fig13_anisotropic_comparison}Comparison of visual inspection and statistical functions. Orange rectangle shows the reproduction of hard data in the input. The calculation of statistical functions is only along southeast direction.}
\end{figure*}

By contrast, the two reconstructions of CGAN (Fig. \ref{fig:Fig12_sandstone_comparison} b and Fig. \ref{fig:Fig12_sandstone_comparison} c) are both visually and quantitatively in accordance with target, unreal structures could be hardly recognized. Also, these reconstructions present seamless transition between the edge of hard data and unknown data. Besides, reconstruction of CGAN only takes $\sim0.08s$, while DS takes $\sim 1.6s$ on the same CPU, achieving a $\sim 20$ speedup factor.

\subsubsection{\label{sec:anisotropic_reconstruction}Anisotropic reconstruction}

We further apply our method to reconstruction of an anisotropic porous material \cite{bostanabad2016characterization}, as demonstrated in Fig. \ref{fig:Fig13_anisotropic_comparison} e. The particularity of this medium is the pore along southeast is significantly longer than other directions. Clearly, even though $128 \times 10$ hard data is given (Fig. \ref{fig:Fig13_anisotropic_comparison} a), the reconstructions are still realistic at visual level, and the hard data is also well preserved. The reconstruction porosity distribution is ${\phi _{CGAN}} = 0.379 \pm 0.064$, which is close to the target ${\phi _{target}} = 0.393$.

As can be seen in Fig. \ref{fig:Fig13_anisotropic_comparison} f-h, the averaged $S_2$ and $L$ of reconstructions are in good agreement with those of target, whereas higher-order function $C_2$ exhibits slight bias, which especially enlarges with the increase of the distance between two pixels. Remarkably, the performance of the proposed method on this anisotropic reconstruction is slightly weak, compared with the three isotropic reconstructions aforementioned. This is primarily due to this material is more topologically complicated in pore¡¯s lineal size and morphology. Additionally, notice that here no explicit prior information associated with direction and size of pore is incorporated during reconstruction, and based on the loss functions used, our method can still capture and reproduce the essential directional morphology and dimensional characteristics. Of course, more constraints can be incorporated into our method to further improve the reconstruction accuracy.

\section{\label{sec:conclusion}Conclusion}

Reconstruction of porous media from limited information has been an outstanding challenge, especially the given data is scarce. To address this problem, in this paper we have presented a deep learning-based method for reconstructing full images of porous media from its much smaller sub-area(s), which provides a framework to produce accurate, fast, and robust realizations. Especially, CGAN is employed to learn the mapping between partial image and full image. In particular, two object functions are proposed and along with other two constraint functions, then jointly constitute the total object function to ensure the reconstruction accuracy. Besides, a Gaussian noise is introduced to preserve reconstruction diversity, allowing for multiple choices for users. Extensively tested on various porous materials, our method has been demonstrated to be able to accurately and stably reconstruct statistically equivalent structures while keeping high efficiency. By using our approach, despite the variety of amount and form of given data, the mapping between input and output can be successfully learned, and consequently the corresponding full images could be instantaneously reproduced. This may be especially useful in the applications that data is lacking or data acquisition is costly.

We would like to highlight that the proposed framework can be readily extended to other applications, such as 2D to 3D or 3D to 3D conditional reconstruction. Theoretically, it is able to incorporate an arbitrary number of object functions of any types into reconstruction, and incorporate any user-define conditional data, which could be of particular use in practice when the presence of specific structures in the area is informed. In addition, it also has the potential to reduce computational cost, for instance, it can be coupled with a variety of reconstruction methods of porous media like MPS \cite{okabe2005pore,gao2015reconstruction,ding2018improved}, DS \cite{mariethoz2010direct}, CCSIM \cite{tahmasebi2012reconstruction,tahmasebi2013cross,tahmasebi2016enhancing,tahmasebi2016enhancinga,tahmasebi2017hypps}, and used for accelerating the matching process in these algorithms. Some of which will be reported in the near future.

\begin{acknowledgments}
This work is supported by the National Natural Science Foundation of China (Grant No. $61372174$). The authors would like to thank Dr. Daniel W. Apley, Department of Industrial Engineering and Management Sciences, Northwestern University, for providing several images of porous media used in this work.
\end{acknowledgments}
\bibliography{References}

\end{document}